\documentclass[preprint2]{aastex}
\slugcomment{}

\shorttitle{Spallation in NGC\,4051}
\shortauthors{Turner et al.}
\def\suzaku{{\em Suzaku}\ }

\begin{document}


\title{
%{\em Suzaku} X-ray spectroscopy of NGC\,4051:
%evidence for cosmic ray spallation 
%and support for the origin of UHECRs in low luminosity AGN.}
Cosmic Ray Spallation in Radio-Quiet Active Galactic Nuclei: A Case Study of NGC 4051}
 
\author{T.J.Turner}
\affil{Department of Physics, University of Maryland Baltimore County, 
   Baltimore, MD 21250 and Astrophysics Science Division,   
NASA/GSFC, Greenbelt, MD 20771, U.S.A}

\and

\author{L.Miller} 
\affil{Department of Physics, University of Oxford, 
Denys Wilkinson Building, Keble Road, Oxford OX1 3RH, U.K.}

\begin{abstract}

We investigate conditions for and consequences of spallation 
in radio-quiet Seyfert galaxies. The work is motivated by the 
recent discovery of significant line emission 
at 5.44\,keV  in \suzaku data from  NGC~4051. 
The energy of the new line suggests an
identification as Cr\,{\sc i}  K$\alpha$ emission, however 
the line is much stronger than would be expected 
from material with cosmic abundances, leading to a suggestion of enhancement owing to 
nuclear spallation of Fe by low energy cosmic rays from the active nucleus. We find that the highest 
abundance enhancements are likely to take 
place in gas out of the plane of the accretion disk and that  timescales for spallation
could be as short as a few years.  The suggestion of a strong nuclear flux of cosmic rays 
in a  radio-quiet Seyfert galaxy is of particular interest in light of the recent 
suggestion from {\it Pierre Auger Observatory} data that ultra-high-energy cosmic rays may originate in 
such sources.

\end{abstract}


\section{Introduction}
\label{introduction}

Recent X-ray data from {\it XMM-Newton}, {\it Chandra} and 
{\it  Suzaku} has led to the discovery of narrow line emission in the 5 -
6 keV regime in several local Seyfert galaxies
\citep[e.g.][]{turner02a, yaqoob03a}.  One popular interpretation of
the lines has been as emission from  so-called 'hotspots' on the
accretion disk, i.e. enhanced emission related to events such as
magnetic reconnections on the disk surface.  The observed line profile 
is modified by Doppler and gravitational effects, depending on the 
inclination of the system to the observers line-of-sight 
and the radial location of the hotspot.   A
hotspot that co-rotates with the disk should
show a periodic pattern of variations in line energy
and strength with time (so long as the disk is not observed face-on).  
If the hotspot originates within 20$r_g$ then
general relativistic effects are predicted to be measurable using
current X-ray data.  Superposed on the periodic shifts,  
lines  should show 
down-shifting in the peak energy of the line 
with time as the material spirals inward \citep{dovciak04a}.
As disk hotspots are not expected to survive for more than a few 
orbits at small radii \citep{karas01a}, observation of persistent lines
apparently originating close to the event horizon would disfavor the 
disk hotspot  model. 

Alternatively, the shifted lines may  arise in 
ejected blobs of gas 
\citep[such as comprising a wind, e.g.][]{turner04a}. 
In contrast to the hotspot model, lines 
originating from  ejecta are expected to show non-periodic 
evolution with time as 
the gas moves outwards, and by tracing the line energy over time 
 one can potentially  constrain the 
 acceleration or deceleration mechanism. 
 In principle then, one could
distinguish disk hotspot and ejecta origins for line emission in the 
5 - 6\, keV band using time-resolved spectroscopy. 

However, another model has been suggested, where observed lines may be 
identified as species of elements such as Cr and Mn that 
normally would be too weak to measure using current data, but whose 
strength has been enhanced owing to abundance changes in the emitting gas 
from spallation of Fe \citep[e.g.][]{turner02a}. 
The interaction of protons having kinetic energy $\ga 30$\,MeV with
matter can result in spallation of its heavy nuclei, creating enhanced
abundances of elements lower in mass than the target nucleus.  
In the low energy regime the protons required for effective spallation
need be only mildly relativistic, easily achievable in a number of
astrophysical situations and therefore we might expect to observe
spallation under a range of conditions. Indeed, spallation is known to
significantly affect the abundance ratios in our own Galaxy
\citep{reeves74a,lund89a} where there is approximate energy
equipartition between cosmic-ray protons and the magnetic and
radiation fields.  The most noticeable effects of spallation are that
abundant nuclei such as C, N, O, Fe are broken down, increasing the
fraction of lighter elements such that the emission or absorption
profile of the gas is significantly different than expected for cosmic
abundance material.  However, the cross-sections are relatively low
($\sigma < 1000$\,mb, \citealt{letaw83a,silberberg98a,tripathi99a},
where 1000\,mb is $10^{-24}$\,cm$^{2}$), and significant abundance
changes in a large mass of gas require either high proton flux and/or
long timescales.

The conditions for and the consequences of cosmic ray production in AGN 
have been discussed in the past by several authors
\citep{kazanas86a,axford94a,cronin05a,nagano2000a}.  
While the effects of spallation could be detectable in any waveband,  the cross-section for spallation
increases with atomic mass approximately as $A^{0.7}$ \citep{letaw83a}; 
this, the high cosmic abundance of Fe and its observability over a
wide range of ionization mean that signatures of spallation may be most easily 
detected in X-ray spectra of iron group elements.
At low cosmic ray energies the primary products from the spallation of Fe, and their
partial cross-sections for prompt production following collision with a 
100\,MeV proton are Mn (400\,mb), Cr (352\,mb), V (115\,mb) and Ti (70\,mb),
where the cross-sections have been determined from code supplied by
\citet{silberberg98a}.  If we adjust the cross-sections to account for the decay
of unstable isotopes, the effective cross-sections become
Mn (247\,mb), Cr (560\,mb), V (163\,mb) and Ti (136\,mb), adopting the
decay channels summarized in Table\,1 of \citet{skibo97} but neglecting
decay of $^{53}$Mn, whose half-life is $3.7 \times 10^6$\,years.
At X-ray energies, neutral species produce K$\alpha$
emission at 5.9, 5.4, 4.9 and 4.5\,keV respectively.
Comparison of the partial cross-sections  shows that the 
most notable  abundance enhancements  produced from 
spallation of Fe would be for Cr and Mn. 

Spallation of Fe has been discussed previously to explain line
emission at 5.6 and 6.1 keV as emission from ionized species of Cr and
Mn in NGC 3516, although in that case the lines ratios did not agree
well with those predicted by \citet{skibo97} for spallation of disk
gas \citep{turner02a}. However, the simulations by \citet{skibo97}
were conservative regarding the level of abundance enhancement that
could occur as a result of spallation in active 
nuclei. \citet{skibo97} considered the requirements for abundance
enhancement in the total mass of material surrounding an accreting
black hole, requiring a high efficiency of proton creation to achieve
high factors of enhancement for Cr and Mn abundance.

In recent years our understanding of the nuclear environs of Seyfert
galaxies has evolved, with ever-increasing evidence for large columns
of gas covering a wide range of ionization state shrouding the active
nucleus and shaping the observed properties of AGN in the X-ray
bandpass.  In this paper we argue that, 
if the gas in the putative disk wind provides the target
for proton spallation, rather than the body of accretion disk itself,
then the predictions and consequences of the process could be very 
different from the calculations of \citet{skibo97}.  

Recent {\it Suzaku} observations of NGC 4051 reveal a line at 
5.44\, keV that is steady in flux and energy over years \citep{turner10a}   
disfavoring  disk hotspot and ejecta models for the line origin and motivating 
a reconsideration of  the importance of spallation in AGN. 
Exploration of the 
conditions for and consequences of spallation in AGN is  the topic 
of this paper.

\section{Summary of Observational Results from NGC 4051}
\label{sec:summary}

Line emission has recently been found at 5.44\, keV and 5.95\, keV  in 2005
and 2008 {\it Suzaku} observations of NGC~4051 \citep{turner10a}.
While potentially of interest with regard to an identification as Mn emission, 
the line at 5.95\,keV
line suffers a degree of contamination from the slightly broadened
component of Fe K$\alpha$ emission \citep{miller09b,turner10a} and (to
a lesser degree) the detector calibration source. Detailed analysis also 
showed the inferred significance of the line at 5.95\, keV to be sensitive 
to the form of the continuum model.  For these
reasons we concentrate on the line at 5.44\, keV and the 
inferred ratio of Fe/Cr in this work. 

Principal components analysis can be used
to decompose data into a mathematical solution showing the
simplest constant and variable orthogonal vector set that can explain
the observations.  We applied this technique to the combined 
2005 and 2008 data from {\it Suzaku} observations of NGC~4051. 
That analysis showed that a line at 5.44\, keV and the
neutral component of Fe K$\alpha$ emission comprise part of a flat
'offset' spectral component that dominates the source spectrum at low
flux levels \citep{miller09b}. 

The statistical case for a line at 5.44\, keV is very strong in NGC~4051. 
The improvement to the spectral fit using a Gaussian line component 
to model the emission is  $\Delta \chi^2= 32$ corresponding to a detection at 
$>99.9\%$ confidence; this was obtained using a local 
continuum parameterization and fitting  the 2005 {\it Suzaku} data. Instead of using a local continuum parameterization one can derive a  
complex model for the full-band {\it Suzaku} and
HETG data and again test for the presence of an additional line. We found that 
such an approach does not diminish the level of confidence at which this 
line  is detected \citep{turner10a,lobban10a}.  
The fit yields a line energy (in the rest-frame of the host galaxy) 
$E=5.44 \pm0.03$\,keV,  line flux  
$n=5.03^{+2.02}_{-2.01} \times 10^{-6}$\,photons\,cm$^{-2}{\rm s^{-1}}$ 
($1 \sigma$ uncertainties are quoted 
throughout) and equivalent width (against the total continuum) 
$46 \pm 16$ eV for 2005 {\it Suzaku} data. In that fit the flux and 
equivalent width of the
 Fe K$\alpha$ line component (observed at an energy $E=6.410\pm0.015$ keV) were 
$n=1.59 \pm 0.23 \times 10^{-5}$\,photons\,cm$^{-2}{\rm s^{-1}}$
and $195 \pm 24$ eV respectively. 
As the width of the 5.44 keV line was not determinable with existing data, 
it was held at 
$\sigma=50$\, eV as determined for 
Fe K$\alpha$  (whose fitted width was 
$\sigma=50^{+26}_{-33}$ eV) and justified based on the 
evidence from principal components analysis 
that the Fe K$\alpha$ emission and the 5.44\, keV emission 
have a common origin \citep{miller09b}. 

The {\it Suzaku} data were  time-sliced and the line
found to be detected in the six resulting time periods, conclusively
demonstrating that it cannot be a statistical fluctuation in the mean
spectrum.  Further to this,  the feature
was determined to be present in all three operational CCDs
\citep{turner10a}.

Finally, we note that the background count rates contributes 
only 2.5\% of the total count 
rate in the 5-7 keV band  for the summed XIS 
spectral data  during the 2005 observation when the source is relatively dim (and 
only 1.3\% when the source is brighter, during 2008).  
There is no evidence for line emission at  5.44\,keV 
in the background spectrum and the combined 
background spectra  from 2005-2008 yielded an upper limit (at 90\% confidence) 
on the flux of such a line to be 
$n < 4.14 \times 10^{-8}$ photons cm$^{-2}{\rm s^{-1}}$, i.e. $< 1$\% of the detected line flux. 
The lack of a feature at comparable 
flux or equivalent width in the background data  also rules out an origin of the 5.44\,keV line 
as a detector feature.

\begin{figure}
\epsscale{1.0}
\plotone{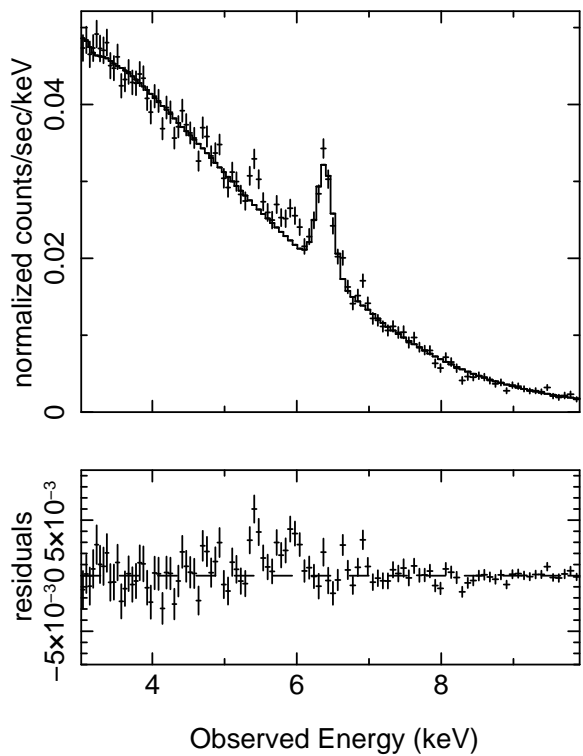}
\caption{\suzaku XIS$0+2+3$ data and residuals from the mean 2005 spectrum, 
compared to an absorbed powerlaw plus Gaussian line at 6.4\,keV, 
showing the residual excess counts 
at $\sim 5.44$ and $5.95$\,keV \label{fig:cts_res}}
\end{figure}

The flux of the line at 5.44\, keV was found to be 
consistent with a constant value across the time-sliced 
{\it Suzaku} data and in archived {\it XMM} data \citep{turner10a};  
 thus the line  shows up most
prominently when the continuum level is low
(Figure~\ref{fig:cts_res}), supporting the result from principal 
components analysis. The line energy was also found to be 
consistent with a constant value and, as noted previously, 
these observations   thus 
disfavor hotspot and ejecta models for the line.  

In the rest-frame of the host galaxy the line may be identified as 
K$\alpha$ emission from Cr {\sc i}, in which case the observed line strength 
exceeds what might be expected from simple continuum illumination of
material with cosmic abundance ratios.  
Strong Cr  emission that appears linked to the Fe emission is 
naturally explained by the spallation process.
The \citet{anders89a} abundance  ratio is 
%200:1:2 between Fe, Mn and Cr;  
100:1 between Fe:Cr; 
taking the fluorescence yields into account 
($Y_{Fe}=0.347$, 
%$Y_{Mn}=0.314$, 
$Y_{Cr}=0.282$) we would expect 
line equivalent widths relative to the illuminating continuum to be observed 
in the approximate ratio 
%246:1:2 for Fe, Mn, Cr 
123:1 for Fe:Cr  
from photoionized gas; folding in the 
relative K-shell photoionization cross sections modifies the estimate to 
an approximate ratio 
%225:1:2 for Fe, Mn, Cr. 
113:1 for Fe:Cr. 
In the absence of spallation, Cr K$\alpha$ line emission 
should be immeasurably weak for most 
AGN observed with current X-ray instruments. 
Contrary to expectations, the observed equivalent width 
ratio in the low-state data for NGC 4051 is Fe/Cr$=4.24\pm1.57$, i.e. 
showing a significant deviation from the expected ratio 
with a factor of  $\sim 30$ enhancement of Cr K$\alpha$ emission 
relative to the line ratios expected  
for cosmic-abundance material. 
The degree of spallation indicated by the ratio Fe/Cr is much
more pronounced than observed in the Milky Way, but   more extreme 
spallation might be expected 
as  very different conditions of cosmic ray flux and material exposure
are thought likely to exist close to an active nucleus. 

Fitting a simple Gaussian model to the  Fe K$\alpha$ emission in NGC~4051 
gave constraints on the width of the line that 
suggested an origin for the emission lines at $r \simeq 0.065$\,light
days, or $2 \times 10^{14}$\,cm, with the 90\% confidence range being
$7 \times 10^{13} - 3 \times 10^{15}$\,cm.
The strength of the Fe line
can be used to constrain the column density of the emitting region
\citep{yaqoob09a} yielding $N_H \ga 10^{24}{\rm cm^{-2}}$, and in the
toroidal reprocessor model suggests a global covering factor 
$C_g \simeq 0.9$.  Detailed spectroscopy of the 2008 {\it Chandra
  HETG} and 2005-2008 {\it Suzaku} data  also show a 
large amount of absorbing gas along the line-of-sight, covering a
range of column densities, ionization states and outflow velocities,
supporting the high global covering derived from the line strength. 
Combining the observational constraints thus yielded a mass
estimate $M \ga 4 \times 10^{-4}$M$_{\odot}$ for the line-emitting gas
with 90\% confidence range $5 \times 10^{-5} < M < 10^{-1}$\,M$_\odot$
\citep{turner10a}.  

While the true total profile of Fe K$\alpha$ may be more complex
\citep{miller09b}, the limit estimated from the simple Gaussian 
model fit
to that line provides a useful mass estimate for examination of the
general feasibility of processes in this case study.

\section{Discussion}

With the evidence for a strong anomalous line in NGC 4051 and the
possible identification as Cr emission we explore the possible
production of cosmic rays in radio-quiet AGN, the dependence of
spallation on parameters of the system, and the consequences of the
process with respect to current observational constraints.

\subsection{Production of cosmic rays in AGN} 

The high luminosity observed in the X-ray bandpass for AGN is thought
to be produced from ultraviolet photons originating in the inner
accretion disk that are up-scattered to X-ray energies
\citep[e.g.][]{haardt91a}.  The requirement for an up-scattering
mechanism has led to the suggestion that a large flux of hot electrons
exist in the vicinity of the inner accretion disk and consequently an
accompanying flux of protons or heavier nuclei may be assumed to be present.  For a
steep power-law cosmic ray energy spectrum most spallation is caused
by low energy protons, $E \la 200$\,MeV, which are only mildly
relativistic ($\beta \la 0.6$).  Shock acceleration could produce such
a population of protons, or mechanisms such as that of
\citet{blandford82a}, where plasma is accelerated by a magnetic field
tied to the rotating accretion disk, which might then generate protons
with those energies.

\subsection{Calculation of abundance enhancements in the ``thick target'' limit}
\label{sec:thicktarget} 
\citet{crosas96a} and \citet{skibo97} have previously considered the possibility of spallation 
in AGN.  \citet{skibo97} considered specifically the spallation of Fe and its detectability
in X-ray spectra, and in this section we follow closely his analysis.
\citeauthor{skibo97} assumed the case where gas presents a thick
target to protons: i.e.  all the cosmic ray energy is absorbed or radiated and there is no
diffusion of cosmic rays out of the region.  
At low proton energies their primary energy loss is
through Coulomb collisions, with an energy decay length, expressed in units
of column density, of approximately
$\Lambda_C \simeq 10^{26}(E/{\rm GeV})^{1.455}\beta $\,atoms\,cm$^{-2}$ 
\citep{skibo97} for protons of energy $E$ and velocity relative to light
$\beta $ .  Thus zones of gas that are opaque to X-ray photons, with
$N_H > 10^{24}$\,cm$^{-2}$, also efficiently absorb cosmic ray protons
with $E \la 100$\,MeV, and for steep incident power-law cosmic ray spectra
the ``thick target'' approximation 
is valid for column densities $N \ga 10^{25}$\,cm$^{-2}$ (the validity
of the thick target approximation is discussed further in 
Section\,\ref{sec:thintarget}).
Within the target, proton directions are efficiently
made isotropic by Coulomb scattering and magnetic field deflections.  
The propagation of cosmic rays through such a target can be treated as
a diffusion problem.  Those cosmic rays then impact heavy nuclei, and the
rate per target nucleus of spallation of species $i$ into species $j$ is 
\begin{equation}
R_{ij} = 4\pi\int_0^\infty\sigma_{ij}(E)J(E){\mathrm d}E,
\label{eqn:rate}
\end{equation}
where $\sigma_{ij}(E)$ is the spallation partial cross-section for that reaction
and $J(E)$ is the intensity in cosmic rays of kinetic energy $E$ 
inside the diffusion region.
In a thick target, $J(E)$ depends on the cosmic ray injection rate
and inversely on the mass $M$ of material through which the cosmic rays
diffuse \citep{skibo97}.
The expectation number of spallation events per target nucleus,
$\langle n_{ij}\rangle$, can then be written as 
\begin{equation}
\langle n_{ij} \rangle = \frac{ \Omega \tau L_{\rm\scriptscriptstyle CR}}{4 \pi M}
f(E_{\rm min},E_{\rm max},\Gamma,\sigma_{ij},\Lambda_C,\Lambda_{\rm inelastic})
\label{eqn:1}
\end{equation}
where $\tau$ is the length of time the material is exposed to cosmic rays,
$L_{\rm\scriptscriptstyle CR}$ is the rate of AGN cosmic ray kinetic energy output and $\Omega$
is the solid angle covered by the target gas. $f$ is a function integrated over
cosmic ray energy whose kernel is a function of the input
spectrum, here parameterized by a powerlaw in kinetic energy of index $\Gamma$
with lower and upper cutoffs $E_{\rm min}, E_{\rm max}$, the spallation 
cross-section and the path lengths due to Coulomb and inelastic collisions,
$\Lambda_{\rm inelastic}$.
\citeauthor{skibo97} equated $\tau$ to the expected time for gas of mass
$M$ to be accreted into the black hole if the mean accretion rate is
$\dot{M}$: $\tau \simeq M/\dot{M}$.  He also wrote 
$L_{\rm\scriptscriptstyle CR} = \eta(r_{\rm\scriptscriptstyle ISCO}) \dot{M} c^2$, 
where $\eta(r_{\rm\scriptscriptstyle ISCO})$ is the efficiency of conversion
of gravitational energy into cosmic ray kinetic energy for accretion of
matter into the innermost stable orbit, from which 
equation\,\ref{eqn:1} may be written as
\begin{eqnarray}
\lefteqn{\langle n_{ij} \rangle = } \nonumber \\
& & \hspace*{-5mm} \frac{ \Omega \eta(r_{\rm\scriptscriptstyle ISCO})c^2}{4 \pi}
f(E_{\rm min},E_{\rm max},\Gamma,\sigma_{ij},\Lambda_C,\Lambda_{\rm inelastic}) \nonumber \\
\label{eqn:2}
\end{eqnarray}
with the mass terms conveniently disappearing.  After taking into account the network of
reactions, \citeauthor{skibo97} found significant
abundance enhancements were achieved for $\eta(r_{\rm\scriptscriptstyle ISCO}) \ga 0.05$,
assuming $\Omega = 4\pi$, and that 
enhancements of Cr comparable to those inferred in section\,\ref{sec:summary} 
were found for $\eta(r_{\rm\scriptscriptstyle ISCO}) \simeq 0.1$.
The \citet{skibo97} assumptions result in a maximum predicted enhancement by a  
factor 15 for Cr K$\alpha$ emission over that expected from unspallated 
solar abundance material, while depleting 
Fe by a factor 2. This enhancement corresponds to a 
 predicted ratio for the line equivalent widths  
 Fe/Cr=3, and  the observed ratio is Fe/Cr=$4.24\pm1.57 $, consistent with this value. 

To achieve consistency with the observed line ratio, the necessary
value for $\eta$ is uncomfortably large. However, the value of $\eta$
is strongly dependent on the assumptions made and hence the timescale
needed for significant spallation; thus we proceed to explore whether
different assumptions may yield consistency with observed data 
{\it  without} the need for very high efficiency or very long timescales
for material bombardment.

\subsection{Spallation in the ``thin target'' limit}
\label{sec:thintarget}
A key assumption in the above is that all the cosmic ray energy is absorbed or radiated 
by a ``thick target''.  However, the column densities required for this assumption to be valid
are $N \ga 10^{25}$\,cm$^{-2}$ at low cosmic ray energies, so we should consider the effect of
loss of cosmic rays from the bombarded region in targets of lower column density.  
In the thick target limit the rate of spallation per target nucleus increases with decreasing
target column density (or mass), and hence for sufficiently low target mass the predicted
spallation rate becomes unphysically large, owing to the neglect of the diffusion of cosmic
rays out of the target.  To solve for the spallation rate while accounting for diffusion losses
would require knowledge of the target geometry and magnetic field strength and structure.  However, we
may {\em estimate} the column density at which the thick-target approximation breaks down by comparing
the calculated collision rate per nucleus with that expected in the thin-target limit,
where a nucleus is exposed to the incident flux of cosmic rays.  Higher energy cosmic rays have
lower Coulomb losses and diffuse more easily from the target, so we should consider the
cross-section for collisions that convert Fe into Cr, as we are interested in the validity
of the approximation for the purposes of calculating Cr abundance enhancement.  This approach
to estimating the limits of validity of the thick target approximation is rather crude, and it neglects
the possibility that a target with a very large internal energy in magnetic field could 
hold an enhanced cosmic ray density compared with its surroundings, but it does give an indication
of the typical column density at which the thick target approximation is likely to break down.

Calculation of the partial cross-section for the spallation of Fe into Cr is described by
\citet{silberberg98a} and we use their code for that calculation.  However, many of
the products of an Fe-p collision are unstable nuclei, so here we follow \citet{skibo97} and
modify the cross-sections to allow for decay into other species 
(see also Section\,\ref{introduction}).  A significant contribution to the net Cr production
arises from decay of $^{50, 52, 54}$Mn (we again neglect decay of $^{53}$Mn owing to its
long half-life).  Including this cross-section in the integration 
kernel of equation\,\ref{eqn:rate}, we find a spallation rate per Fe nucleus of
$R_{\rm Fe \rightarrow Cr} \simeq 2 \times 10^{-8}(10^{24}{\rm cm}^{-2}/N_{\rm H})$\,s$^{-1}$,
assuming a cosmic ray spectral index $\Gamma=2.4$, a low energy cut-off 
 10\,MeV and a 
cosmic ray luminosity of $10^{43}$\,erg\,s$^{-1}$
irradiating a target shell at distance $2 \times 10^{14}$\,cm.
We may compare this with the rate expected in the thin target limit, for which we find 
$R_{\rm Fe \rightarrow Cr} \simeq 3.8 \times 10^{-8}$\,s$^{-1}$, so we expect the transition from
the thick to thin target regimes to be important at column densities $N_{\rm H} \simeq 10^{24}$\,cm$^{-2}$,
close to the value we infer for the line-emitting gas in NGC\,4051.  For lower column densities
it is essential to use the more conservative thin target limit.

In terms of the efficiency of
spallation, the thick target limit ensures a maximum number of spallations occur per cosmic
ray, and hence the maximum amount of enhanced material is produced.  
The total amount of enhanced material would be less in the thin target case, 
but the number of spallations per target nucleus reaches a maximum, so 
the abundance change in the gas being bombarded could be more marked in a target of
lower column density.
We return to consideration of spallation efficiencies in section\,\ref{sec:location}.

\subsection{Efficiencies, masses and timescales}
A crucial parameter in the calculation of abundance change is the timescale over which nuclei
are exposed to cosmic rays.
From the above we can see that a high value of $\langle n_{ij} \rangle$ may be obtained either
by a high cosmic ray flux or by a long timescale, greater than $M/\dot{M}$.  
Assuming a bolometric luminosity $L_{bol}= 10^{43} {\rm erg\, s^{-1}}$ 
(\citealt{vasudevan09a}, corrected to the Tully-Fisher distance 15.2\,Mpc, \citealt{russell04a}) and
accretion bolometric radiative efficiency $\eta^{\rm BOL} = 0.05$ we deduce a mass accretion rate 
$\dot{M} \simeq 0.0035$\,M$_\odot$year$^{-1}$.  The mass of gas bombarded by the cosmic
rays is model-dependent, as we discuss below, but for a generic mass of $M \sim 0.01$\,M$_\odot$
the accretion timescale $M/\dot{M}$ is only about 3\,years.  
This timescale is much less than the current
growth timescale of the black hole, $\tau_{BH} \simeq M_{BH}/\dot{M} \simeq 5 \times 10^8$\,years
for black hole mass $M_{BH} =  1.7 \times 10^6 M_{\rm \odot}$ \citep{denney09a}
and leaves open the possibility that gas could achieve significant enhancements in abundance
of  Cr with a low cosmic ray luminosity, if it can survive near the black hole without
being either accreted or blown out.  

\subsection{Location of the enhanced gas}
\label{sec:location}
Arguably the first location to consider for the target gas is the
accretion disk.  However, simple models of spallation within the disk
both require the spallation timescale to be sufficiently long to
generate the inferred abundance changes, $\tau \simeq M_{\rm
  disk}/\dot{M}$, and predominantly generate enhanced abundances only
in the inner regions of the disk (see Appendix).  Spallation may well
occur within the accretion disk, but it seems unlikely that the
efficiency would be high enough to achieve the abundance changes that
we observe, and we would require a mechanism that transports enhanced
material from near the ISCO to larger radii for consistency with the
line widths that we observe.  While transportation of the spallated
gas is possible via outflow mechanisms such as a disk wind, the
spallated gas detected in NGC\,4051 yields a limit on the outflow bulk
velocity $v < 1200$ km s$^{-1}$. It is difficult to determine from
X-ray spectroscopy whether the component of outflowing gas traced by
He-like and H-like Fe absorption lines has been modified by
spallation, as Cr would be highly ionized in such a zone and
would provide little X-ray opacity.

A more attractive possibility is that the target gas is not part of
the accretion disk itself, but rather is material lying out of the
plane of the disk, perhaps even in or near the broad line region.  In
this case a relatively small mass of gas could intercept a large
fraction of the cosmic rays emitted from the AGN, leading to much
higher spallation rates than would be inferred for spallation within
the accretion disk.  Detailed timing analysis of the NGC~4051 data 
shows frequency-dependent 
time lags between the hard and soft X-ray photons of up to 970s. 
These may be explained by the effect of reverberation in the hard X-ray band 
as continuum photons are reflected from a 
thick shell of circumnuclear material extending to $\sim 1.5 \times 10^{14}$ cm 
from the black hole, having global covering factor $\ga 0.44$ \citep{miller09b}. 
This shell could comprise at least some fraction of the target gas. 

To estimate the expected spallation timescale in the thick target
approximation, we should assume a high column density, $N \ga
10^{25}$\,cm$^{-2}$, and equivalently a relatively high mass of target
gas, $M_{\rm gas} \ga 4 \times 10^{-4}$\,M$_\odot$, for the nominal
radius of $2 \times 10^{14}$\,cm and high filling factor
(section\,\ref{sec:summary}; recall that in the thick target
approximation, the abundance enhancement for a given cosmic ray
exposure time decreases with $M$).  Comparing with the \citet{skibo97}
calculation, the exposure to cosmic rays would need to last $\tau \ga
M_{\rm gas}/\dot{M} \ga 4 \times 10^{-4}/0.0035 \simeq 0.11$\,years
for $\eta=0.1$ (equivalent to a cosmic ray luminosity $2\times 10^{43}$\,erg\,s$^{-1}$).  
If we adopt a less extreme value for the efficiency
of cosmic ray production, we would still only require the gas to be
exposed to cosmic rays for a time $\tau \ga 1.1/\eta$\,years.  The
reason for the dramatic change in timescale, and the less restrictive
requirement for the efficiency $\eta$, compared with the accretion
disk case or with the \citet{skibo97} analysis, is the higher rate of
cosmic ray bombardment per Fe atom in the target gas.  A small
fraction of the gas in the accreting system receives a
disproportionately high fraction of the cosmic rays.  This of course
is quite natural in the standard accretion disk picture, where most of
the mass is in the disk but where additional traces of material are
exposed to the central source over the remainder of the $4\pi$\,sr.

For lower column densities and target masses, the timescale needed for
significant spallation is expected to asymptotically approach the
inverse spallation rate $\tau \rightarrow 1/\dot{R}$ for exposure to
the incident cosmic ray flux (section\,\ref{sec:thintarget}), which
for Fe with the assumed cosmic ray spectrum, luminosity $10^{43}$\,erg\,s$^{-1}$
and radius $2\times 10^{14}$\,cm is
$\tau \ga 0.9$\,years. This estimate for low column densities
yields a similar value to the thick-target calculation, 
albeit not one that takes into account the network of reactions as in 
the calculation of \citet{skibo97}.

If we knew the detailed geometry and magnetic field structure of the target 
we should solve the diffusion-loss
equation for some assumed system mass and diffusion loss coefficient,
but in either the thin- or thick-target case
the required timescale is only of order years, and over the lifetime
of the black hole growth, a substantial total amount of material could
be processed through the cosmic ray region: at any one instant of time
we only need to be observing a small mass of spallation-enhanced gas
which could have been substantially enhanced in its Cr 
abundance.  This scenario is an effective way of explaining the
apparent abundance enhancements that we see without requiring extreme
conditions of cosmic ray production.

\subsection{Gamma-ray and radio emission}
Interactions of cosmic rays with the accreting gas are expected to 
produce secondary $\gamma$-ray emission 
\citep{dermer86a}  and \citet[][equations 11-12]{skibo97} 
estimated the $\gamma$-ray flux above 100 MeV for Seyfert galaxies if spallation is occurring. 
We have recalculated the expected $\gamma$-ray flux specifically for NGC\,4051 following
\citet{skibo97}.  The estimate considers only production of $\gamma$-rays following neutral pion
creation, as calculated for the Milky Way by \citet{dermer86a}, whose code we use for the
calculation, modified to incorporate equation\,4 of \citet{skibo97}.  The calculation assumes
the ``thick target'' case where all protons are absorbed, and hence this provides a maximum
$\gamma$-ray luminosity for a given output AGN proton luminosity.
Assuming that $L_{\rm\scriptscriptstyle CR} = L_{bol}$, the predicted $\gamma$-ray flux is 
$F (> 100\, {\rm MeV}) \simeq 3.3 \times 10^{-8} (\Omega/4\pi){\rm photons\, s^{-1}\, cm^{-2}\,}$  
for proton spectral index $\alpha=2.4$ and proton low-energy cut-off 10\,MeV. 
About 0.8\,percent of the input cosmic ray energy is radiated as $\gamma$-rays.
For this proton input spectrum, the $\gamma$-ray spectrum is harder, 
with an approximately power-law
form with photon index 2.1 in the range 100\,MeV$ < E_\gamma < 100$\,GeV. 
\citet{abdo09a} discuss the detection limit of the {\it Fermi} 
Large Area Telescope and for $\Omega \sim 0.9$ the predicted flux of NGC\,4051 falls just below the 10$\sigma$ 
detection limit from the 3-month sky survey data 
$\sim 3.5 \times 10^{-8}  {\rm photons\, s^{-1}\, cm^{-2}\,} $ 
for a source of this spectrum and Galactic coordinates \citep{abdo09a}.  
A ``thinner'' target in which protons escape the AGN, 
lower global covering factors, 
lower cosmic ray luminosities achieved through longer spallation timescales 
or the possibility of intermittent cosmic ray output 
all weaken the possible constraints available from measurements
of $\gamma$-ray flux. 

Radio synchrotron emission might also be expected, partly as a result of secondary electron creation
but more significantly through primary cosmic ray electrons generated
as part of the process that accelerates the protons.  Any
estimate of the synchrotron emission is extremely uncertain, as we don't know the efficiency of
electron acceleration compared with proton acceleration, we don't know the magnetic field strength
and the electrons themselves are expected to suffer significant ionization losses propagating
through the material as well as adiabatic and radiative losses through bremsstrahlung,
synchrotron and inverse Compton scattering. Furthermore, any compact synchrotron-emitting
region would be self-absorbed, significantly reducing the detect-ability of any synchrotron
emission.  We do not here solve the diffusion-loss equation for electrons, but we can already place
a stringent limit from simple estimates of the radiative losses.
The ratio of synchrotron to inverse Compton losses is given approximately by the ratio of energy
densities in magnetic fields, $u_{mag}$, and radiation, $u_\gamma$, 
so if these were the dominant energy loss mechanisms, 
we would expect an integrated synchrotron luminosity 
\[
L_{\rm sync} \simeq L_{\rm electron} \frac{u_{\rm mag}}{u_\gamma + u_{\rm mag}}
\]
neglecting synchrotron self-absorption for the moment.  If other mechanisms lead to comparable
energy losses, the synchrotron luminosity would be lower.
NGC\,4051 has an steep spectrum radio core with flux
density 3.2\,mJy at 5\,GHz when measured at resolution 1.1\,arcsec \citep{ho01a}, corresponding
to a spatial resolution 80\,pc.  The core has been claimed to be 
marginally detected at EVN resolution \citep{giroletti09a}.  
The core radio luminosity of NGC\,4051
integrated up to 100\,GHz is approximately $3 \times 10^{37}$\,erg\,s$^{-1}$,
so we can place an approximate upper limit on the magnetic field strength if we require
$L_{\rm sync}$ to be less than this value.  For
$L_{\rm electron} \simeq L_{\rm bol} \simeq 10^{43}$\,erg\,s$^{-1}$
and an energy density in radiation at $r=2 \times 10^{14}$\,cm of $u \simeq 700$\,erg\,cm$^{-3}$
we find $u_{\rm mag} < 0.002$\,erg\,cm$^{-3}$, corresponding to a magnetic flux density
$B < 0.2$\,Gauss.  In the synchrotron-optically-thin regime a discrepant radio flux
could only be produced if $B$ were higher than this value, but then
any mJy synchrotron source with magnetic field higher than this value would
be strongly self-absorbed at GHz frequencies on these angular scales.
So even without consideration of the other
energy loss mechanisms, we conclude that no significant excess of radio emission would be
expected from the region.  If some fraction of relativistic electrons escape to larger
radii where inverse Compton losses are less severe and where the synchrotron surface 
brightness limit is not so restrictive, detectable radio emission could be produced, but
without a detailed model for the electron energy losses and electron transport to larger
radii, we cannot estimate its possible contribution to the observed radio emission.

\subsection{Energy and momentum transfer}
At low cosmic ray energies $\la 1$\,GeV, the cosmic ray energy is absorbed by either Coulomb  
interactions or
inelastic scattering and spallation, and ultimately that energy ends up as either heat 
or kinetic energy (see also \citealt{crosas96a}).  For a ``thin'' target the heating may
be relatively low, but for a thick target
we can estimate the energy density in the
diffusing cosmic rays by evaluating the cosmic ray intensity $J(E)$ from equation\,4 of
\citet{skibo97}.  Assuming $L_{\rm\scriptscriptstyle CR} = 10^{43}$\,erg\,s$^{-1}$ we estimate
an energy density 
$u_{\rm\scriptscriptstyle CR} \simeq 240/N_{25}$\,erg\,cm$^{-3}$, where
$N_{25}$ is the column density in units of $10^{25}$cm$^{-2}$, assuming a powerlaw cosmic
ray spectrum with index 2.4 above 10\,MeV and a source-gas distance of $2\times 10^{14}$\,cm. 
The expression is invalid for $N_{25} < 1$ as the gas would no longer satisfy the ``thick
target'' assumption.  For $N_{25} = 1$ the energy density is equivalent to the thermal
pressure in gas of temperature $10^6$\,K with density $10^{12}$\,cm$^{-3}$, so in sharing
their energy with the absorbing gas, the cosmic
rays make a significant contribution to the thermal history of a thick target.  The momentum
transfer is similarly significant: in an equivalent of the ``Eddington luminosity'' calculation,
for $L_{\rm\scriptscriptstyle CR}=10^{43}$\,erg\,s$^{-1}$ and column density $N=10^{26}$\,cm$^{-2}$
the rate of cosmic ray momentum transfer is 0.4 of the Eddington rate.  These high values of
energy and momentum deposition do not apply in the ``thin target'' regime, but if there were
significant column densities of material intercepting the cosmic rays, they would be expected
to experience significant heating and radial forces, as suggested by \citet{sironi09a}.

\subsection{On the origin of UHECRs}
The astrophysical origin of cosmic rays is a topic of great interest.
Cosmic rays up to energies of $10^{15}$ eV appear to originate from
supernova remnants, those in the range $10^{15}<E<10^{18}$ eV also
appear to have a Galactic origin, however, the source of ultra-high
energy cosmic rays (UHECR) $E> 10^{18} $ eV is not yet accounted for.
The extreme conditions in the nuclei of active galaxies offer an
appealing possibility for production of UHECRs but cosmic rays have
not yet been directly established as originating from active nuclei. 
While the cosmic rays required for spallation
are of low energy, $\la 1$\,GeV, the mechanisms thought to 
be in play in AGN  may also produce UHECRs.

To date, radio loud AGN have been the focus of studies 
of cosmic-ray production in AGN because radio  jets have been widely 
thought to provide a suitable site for particle  acceleration.
However, even the jets of radio-loud AGN have fine-tuned requirements for successful 
particle acceleration: 
\citet{farrar09a} argue that shock acceleration in a relativistic jet
is constrained by the need to have a sufficiently strong
magnetic field for acceleration of particles to 
energies above $10^{18}$ eV, while staying within a range that avoids overproduction of 
synchrotron radiation and excessive photo-pion energy losses in the AGN.  

 It had been thought that UHECRs could not be produced in the nuclear
 regions of radio-quiet AGN because of the  small size of the 
 acceleration region and the energy losses
 \citep[e.g.][]{norman95a}.  However new work by \citet{peer09a} has
 argued that nearby radio-quiet AGN could indeed be the source of
 UHECRs. \citet{peer09a} consider particle acceleration in the
 parsec-scale weak jets that are known to exist in many local
 radio-quiet AGN including NGC 4051  \citep{giroletti09a}. 
%Initial calculations had dismissed these low
% luminosity jets as candidate regions for particle acceleration for
% proton UHECRs.  
However, recent {\it AUGER} results have suggested
that UHECRs may be dominated by heavy nuclei \citep{unger07a,bellido09a},
and in this case the constraints on jet luminosity are reduced.
 \citet{peer09a} calculate conditions for a bolometric
 luminosity $10^{43} {\rm erg\, s^{-1}}$, applicable to NGC 4051,
 and find that the nucleus can survive photo-disintegration if the
 acceleration occurs on a parsec scale,  
 concluding that radio-quiet AGN are viable sources of
 UHECRs.

The relationship between the observed photon flux and currently observed cosmic ray flux from a 
particular galaxy is unclear, owing to the significant time delay that must exist 
for the arrival at Earth of cosmic rays
compared with X-ray photons (of order 10$^5$ years; \citealt{moskalenko09a}).  
However one can attempt to conduct a 
statistical survey of the coincidence of UHECR occurrence  and the positions of AGN. 
Intriguingly, \citet{zaw09a}  have discovered a  significant angular
correlation between low luminosity AGN and high-energy cosmic rays in data from
the Pierre Auger Observatory. The correlation is too strong to be simply explained 
by AGN tracing the large-scale distribution of matter, 
indicating that a significant fraction of 
UHECRs are produced in AGN \citep{farrar09b}. 
%The bolometric luminosities of most of the candidate  AGN
%are significantly lower than those required to satisfy the minimum condition for UHECR 
%acceleration in a continuous jet. Thus it appears that UHECRs are not made in the jets of 
%radio loud AGN, but rather in the nuclei of modest luminosity, radio quiet Seyfert galaxies. 

\citet{moskalenko09a} suggest that as the UHECR candidate AGN have no
special characteristics to set them apart from the AGN population (i.e. they are not 
the radio-loud sources) 
then the apparent correlation of low luminosity AGN with UHECRs must
be a chance occurrence.  
\citet{zaw09a} note that the candidate AGN
may not appear 'special' at any given time if cosmic ray acceleration
occurs via a sporadic event such as tidal disruption, which may inject
$> 10^{52}$ ergs into the nuclear environs \citep{farrar09a} and yet not be occurring 
at the time of observation of the candidate AGN. Tidal
disruption events should only occur every $10^4-10^5$ years
\citep{magorrian99a} in a given AGN and  the
duration of the event itself would likely be very brief, of the order
of tens of rotations at the innermost stable orbit, i.e.  hundreds to 
thousands of seconds in the case of NGC 4051.

However, current data cannot rule out the possibility that modest
luminosity, radio quiet AGN commonly and persistently produce cosmic
rays in their nuclear environs.  As the typical number of UHECR
detected is only $\sim 1$ per AGN, Poisson sampling guarantees that
not all AGN will have associated UHECR in the Pierre Auger Observatory
data.  If UHECR production is linked to production of low energy
cosmic rays in AGN, signatures of spallation may be better indicators
of cosmic ray acceleration rather than large-scale radio jets, and
invoking a sporadic event for cosmic-ray acceleration appears 
unnecessary.  If spallation is occurring in NGC 4051, then this provides
important observational evidence for the existence of a significant
flux of low energy cosmic-rays in radio-quiet active 
galactic nuclei, and by extension is
supportive of the production of UHECRs from the same sources. 

\section{Conclusions}

Significant line emission at   5.44\, keV in NGC 4051 
may be interpreted as evidence for spallation at work in the nucleus of a radio-quiet active galaxy. 
Examining conditions for and consequences of  spallation  in circumnuclear material 
we find the highest abundance enhancements are likely to take
place in target gas away from the plane of the accretion disk with timescales for spallation
that could be as short as a few years if the cosmic ray output 
is comparable to the bolometric output of the active galaxy. 
Spallation timescales would be proportionally longer for lower cosmic ray luminosity,
but even so could still be orders of magnitude shorter than the accretion timescale onto the black hole.
Feedback of material from the active nucleus to the host galaxy 
may result in enhanced abundance ratios at larger radii than probed with these X-ray observations.

As the acceleration processes required to produce the protons necessary 
for spallation likely produces a broad spectrum of energetic particles, the results are 
also interesting in the context of our broader understanding of the origin of cosmic rays. 
The evidence for spallation in this radio-quiet active galaxy may support the discovery from the 
{\it Pierre Auger Observatory} data that the origin of ultra-high-energy cosmic rays 
may be the nuclei of radio-quiet AGN rather than luminous radio-loud AGN.

\acknowledgments

We are grateful to C.\,Dermer for supplying 
his code to calculate secondary $\gamma$-ray emission from neutral pion production,
TJT acknowledges NASA grants NNX09AO92G and GO9-0123X. LM acknowledges STFC grant number PP/E001114/1.

\appendix

\section{Appendix material}

\subsection{Accretion disk spallation}

We might suppose that material in the accretion disk is exposed to cosmic rays.  However,
in the simplest pictures, spallation would occur primarily close to the black hole and would
be limited by the accretion timescale assumed by \citet{skibo97}, $\tau \simeq M/\dot{M}$.  
Consider accretion disk material being illuminated by cosmic rays from a source a height $d$ above
the disk on the rotation axis.  An annulus on the disk
containing mass $\mathrm{d}M$ and subtending a solid angle $\mathrm{d}\Omega$ at $d$,
has a cosmic ray intensity
\[
J \propto \frac{L}{4\pi}\frac{\mathrm{d}\Omega}{\mathrm{d}M}
\]
if the cosmic rays are localized to that radius, as expected for the very high column density
in the accretion disk.  If mass flows through that annulus at a radially-invariant rate
$\dot{M}$ the expectation value for the number of spallations per nucleus is
\[
\mathrm{d}\langle n_{ij}\rangle \propto \frac{L}{4\pi}\frac{\mathrm{d}\Omega}{\dot{M}}
\]
and so the total number of spallations per nucleus integrated from infinity to radius
$r$ is
\[
\int_\Omega^{2\pi}\mathrm{d}n(\Omega) = \frac{L}{2\dot{M}}\frac{d}{\sqrt{d^2+r^2}}.
\]
Thus half the expected spallation events occur within a radius $r_{1/2}=\sqrt{3}d$.
The total number of spallations is half the value calculated by \citet{skibo97} owing
to the global covering factor of the disk of 0.5, but the spallation timescale has the 
value he assumed.  

We reach a similar conclusion if we suppose instead that cosmic ray sources, perhaps
associated with shocks in the accretion disk, are embedded within the disk.  We gain
a factor 2 if the cosmic rays are now all intercepted by thick disk material.
If the cosmic ray intensity as a function of radius is given by the gradient in
potential energy, then for Keplerian orbits in a Newtonian potential $\phi$ we expect
\[
J \propto \frac{\epsilon c^2}{2}\frac{\mathrm{d}\phi}{\mathrm{d}t},
\]
where $\epsilon$ is the efficiency of converting gravitational potential energy
into cosmic ray energy\footnote{The relationship between $\epsilon$ and $\eta(r)$
for accretion in to radius $r$ is 
\[
\eta(r) = \frac{\epsilon}{2c^2} \int_r^{r=\infty}\mathrm{d}\phi = \frac{\epsilon G M_{BH}}{2 rc^2}
\]
assuming Keplerian orbits in Newtonian gravity and 
\[
\eta(r) = \epsilon\left[1-\frac{\left(1-2GM_{BH}/rc^2\right)}{\sqrt{1-3GM_{BH}/rc^2}}\right]
\]
for circular orbits in a Schwarzschild metric.
Note that the energy released from accretion must be partitioned between all forms of
energy output from the active nucleus:  i.e. it is not possible to have a Schwarzschild black
hole producing electromagnetic radiation and cosmic rays with $\eta(r_{\rm\scriptscriptstyle ISCO})=0.057$ 
in {\em both} outputs simultaneously.
}.
Note that the mass of an element of gas cancels as in \citealt{skibo97}.
The expected number of spallations per nucleus at radius r is
\begin{equation}
\langle n_{ij} \rangle = \eta(r) c^2
f(E_{\rm min},E_{\rm max},\Gamma,\sigma_{ij},\Lambda_C,\Lambda_{\rm inelastic})
\label{eqn:3}
\end{equation}
which is the relation given by equation\,\ref{eqn:2} 
for $r = r_{\rm\scriptscriptstyle ISCO}$ and $\Omega = 4\pi$.
The value of $\eta(r)$ should be calculated appropriately for some assumed
spin of the black
hole, but for any metric the spallation occurs predominantly close
to the ISCO.

Thus in either case, we reach the follow conclusions: (i) the spallation timescale
is about the same as the accretion timescale, $\tau \simeq M/\dot{M}$, so to achieve
large abundance changes a high efficiency of cosmic ray production is needed,
$\eta(r_{\rm\scriptscriptstyle ISCO}) \ga 0.1$; (ii)
spallation in the accretion disk would be expected to occur
primarily in the inner regions unless the cosmic ray source is very extended and/or
distant from the disk.
Yet the Fe\,K$\alpha$ line, which we suppose originates in the same enhanced gas, 
is narrow, with a likely location
$r > 7 \times 10^{13}$cm (section\,\ref{sec:summary}).  However, we don't necessarily
observe the gas in the same location as where the spallation occurred:
thus an accretion disk origin for the spallation 
would require gas that has been enhanced near the black hole to have been transported back
out to the large radii where we observe it.

\bibliographystyle{apj}      % basic style, author-year citations
\bibliography{xray_nov_2009}   % name your BibTeX data base

\end{document}